\documentclass[aps,prl,twocolumn,nofootinbib,preprintnumbers]{revtex4-1}
\usepackage{amsmath,amssymb, amsfonts}
\usepackage[bookmarks=true,colorlinks,linkcolor=red,urlcolor=blue,citecolor=blue]{hyperref}
\usepackage[usenames]{color}
\usepackage{dcolumn}
\usepackage[normalem]{ulem}

\usepackage{bm}

\usepackage{feynmf}

\usepackage{graphicx}

\newcommand\dr{\mathrm{d}}

\renewcommand{\kappa}{q}

\def\({\left(}
\def\){\right)}

\newcommand\half{{\ensuremath{\frac{1}{2}}}}
\def\eg{{\it e.g.}}
\def\ie{{\it i.e.}}

\def\Dslash{\rlap{\hskip0.2em/}D}
\newcommand\field[1]{{\ensuremath{\mathbb{{#1}}}}}
\newcommand{\IR}{\field{R}}

\newcommand\sD{{\ensuremath{{\mathcal D}}}}
\newcommand\uz{{\underline{z}}}
\newcommand\ut{{\underline{t}}}

\newcommand\bpsi{{\bar \psi}}

\newcommand{\be}{\begin{equation}}
\newcommand{\ee}{\end{equation}}
\newcommand{\bea}{\begin{eqnarray}}
\newcommand{\eea}{\end{eqnarray}}
\newcommand{\bwt}{\begin{widetext}}
\newcommand{\ewt}{\end{widetext}}

\newcommand{\bi}{\begin{itemize}}
\newcommand{\ei}{\end{itemize}}
\newcommand{\ben}{\begin{enumerate}}
\newcommand{\een}{\end{enumerate}}
\newcommand{\bca}{\begin{cases}}
\newcommand{\eca}{\end{cases}}
\newcommand{\bln}{\begin{align}}
\newcommand{\eln}{\end{align}}
\newcommand{\bst}{\begin{split}}
\newcommand{\est}{\end{split}}

\def\yandz{\Psi}
\def\psis{\yandz}

\newcommand\ux{{\underline{x}}}
\newcommand\uy{{\underline{y}}}

\def\Ione{\hbox{$1\hskip -1.2pt\vrule depth 0pt height 1.53ex width 0.7pt
                  \vrule depth 0pt height 0.3pt width 0.12em$}}

\def\vev#1{\langle#1\rangle}

\def\tr{{\rm tr\ }}
\def\Re{{\rm Re\hskip0.1em}}
\def\Im{{\rm Im\hskip0.1em}}

\preprint{MIT-CTP/4342}
\begin{document}
\title{A quantum electron star}
\author{Andrea Allais, John McGreevy and S. Josephine Suh}
\affiliation{Department of Physics, Massachusetts Institute of Technology, Cambridge, MA 02139}
\begin{abstract}
We construct and probe a holographic 
description of 
state of matter which results 
from coupling
a Fermi liquid to a relativistic conformal
field theory (CFT).  
The bulk solution is described by a quantum 
gas of fermions supported 
from collapse into the gravitational well of AdS
by their own electrostatic repulsion.
In the probe limit studied here, the 
Landau quasiparticles survive this coupling to a CFT.
\end{abstract}
\maketitle
\section{Introduction}

Some progress has been made recently 
in using holography to learn about 
metallic states of matter which are not 
described by Landau's Fermi liquid theory 
\cite{Lee:2008xf, Liu:2009dm, Cubrovic:2009ye, Faulkner:2009wj, Faulkner:2011tm, Faulkner:2010zz, 
Faulkner:2010tq, Hartnoll:2009ns, 
Hartnoll:2010gu, Hartnoll:2010xj, Hartnoll:2010ik, Hartnoll:2011dm, Puletti:2010de, Puletti:2011pr}.
In particular, the work of \cite{Liu:2009dm, Faulkner:2009wj}
constructs controlled non-Fermi liquid fixed points,
which however constitute a parametrically-small (in the bulk 
Newton's constant $1/N^2$) fraction of a larger system;
this larger system is responsible for the destruction of the Landau quasiparticle.

The fact that this larger system is locally critical is 
the origin of the short transport lifetime and linear-$T$ resistivity 
in the case which realizes marginal Fermi liquid \cite{Faulkner:2011tm, Faulkner:2010zz}.
This local criticality is further closely tied to the nonzero zero-temperature entropy of this state,
indicating that instabilities prevent one from cooling into this state.
Various instabilities have been suggested;
an intrinsic one, arising from the density of fermions itself,
was pointed out in 
\cite{Hartnoll:2009ns}:
these fermions screen the gauge flux
which supports the $AdS_2$ throat,
and a truer groundstate is a Lifshitz geometry with large dynamical exponent $z \sim N^2$,
rather than the $z=\infty$ of $AdS_2 \times \IR^2$.
Such `electron star' states,
comprising a charged and gravitating density of fermions in the bulk,
have been further studied in detail 
\cite{Hartnoll:2010gu, Hartnoll:2010xj, Hartnoll:2010ik, Hartnoll:2011dm, Puletti:2010de, Puletti:2011pr}.
These states represent an improvement over the work of 
\cite{Liu:2009dm, Faulkner:2009wj} in 
that the fermions play a leading order (in $N^2$) role in constructing the geometry.
However, 
the single-fermion response in these states exhibits
{\it many} Fermi surfaces \cite{Hartnoll:2011dm, Puletti:2011pr},
rather than one.
This conclusion is not intrinsic to holographic states
supported by fermions, but rather is an artifact of the Thomas-Fermi approximation
used in the pioneering work on electron stars \cite{Hartnoll:2009ns, Hartnoll:2010gu, Hartnoll:2010xj, Hartnoll:2010ik, Hartnoll:2011dm, Puletti:2010de, Puletti:2011pr}. 

We would like to construct
a holographic non-Fermi liquid (NFL) where
the fermions contribute at leading order to the state
({\it e.g.}~to the thermodynamics and groundstate entanglement entropy of regions).

A point of departure
is provided by \cite{Sachdev:2011ze}\footnote{Earlier work which studies
quantum spinor fields in a holographic context includes \cite{Cubrovic:2010bf, Cubrovic:2011xm}.  
Recent provocative work towards this goal includes 
\cite{Ogawa:2011bz, Huijse:2011ef, Shaghoulian:2011aa, Dong:2012se}.}.
The state constructed there can be understood as 
a Fermi liquid coupled to a (toy model of a) confining gauge theory.
The outcome is a Fermi liquid, in bulk and boundary.  The lifetimes of the quasiparticles
are infinite in the leading large-$N$ limit.

This paper makes a further step towards the above goal for holographic NFLs.
We construct a bulk Fermi liquid state
without putting in a gap in the geometry from the beginning.
That is, we show that the bulk state constructed in \cite{Sachdev:2011ze}
survives the limit where the hard-wall cutoff is taken away.
The problem we solve is the fermion analog of 
the probe holographic superconductor calculation in
\cite{HHH1}.
The resulting state is a {\it quantum electron star}\footnote
{This usage has appeared in \cite{Hartnoll:2011dm}.}
in the sense that we are treating the bulk fermions quantum mechanically,
but they do not collapse into the Poincar\'e horizon.
Solving this problem is a prerequisite for the 
more general problem including gravitation.

Technically, this required some improvements over previous methods.
In particular, in order to understand accurately the 
effect of the bulk electric field on the 
charge of the filled Dirac sea, it was necessary to 
provide a short-distance (UV) completion of the bulk system.
We accomplish this by putting the bulk system on a lattice.
As a result, we were forced to account for the polarization of the Dirac 
vacuum by the electrostatic potential, 
and the resulting screening of the electric charge.

Next, we describe the problem, and outline our method of solution, deferring details to the appendices.
Results and discussion follow.  
The appendices also include a discussion of 
several interesting effects uncovered here,
including the chiral anomaly in the bulk,
and a surface charge density.

\section{Setup of the problem}

We consider a system defined by the action
\begin{equation}
S =  \int \dr^{d+1} x \sqrt{-g} \left[ \frac{ R - 2 \Lambda}{16\pi G_N} - {1\over 4 q^2} F^2 \right] + W[A]\,,
\end{equation} 
where
\begin{equation}\label{eq:wtf}
 e^{iW[A]} = \int \mathcal{D}\psi\ e^{i S_f[\psi, A]}\,,
\end{equation} 
\begin{equation}
 S_f[\psi,A] = \int \dr^{d+1} x \sqrt{-g}\ \left[- i \bar \psi \left(\Gamma^M {\cal D}_M + m\right) \psi\right]\,,
\end{equation} 
$\bpsi \equiv \psi^\dagger \Gamma^\ut$ and
$
 \sD_M  \equiv \partial_M + {1 \over 4} \omega_{ab M} \Gamma^{ab} + i A_M \,,
$ 
with $\omega_{ab M}$ the spin connection.
 
We study this system in a probe limit, $G_N \to 0$ at fixed $\Lambda$, where the geometry is not dynamical. In the dual language, we are studying a CFT where the fraction of degrees of freedom that carry charge is small. 

In particular, we specialize to the AdS$_4$ metric
\begin{equation}
 \dr s^2 = \frac{1}{z^2} \left(-\dr t^2 + \dr x^2 + \dr y^2 + \dr z^2\right)\,,
\end{equation} 
and we consider a gauge field of the form $A = \Phi(z) \dr t$. The equation of motion for the potential $\Phi$ is
\begin{equation} \label{gauss law 1}
- \Phi''(z) = q^2 z^{-3} \left\langle{\hat \psi}^\dag\hat \psi\right \rangle\,.
\end{equation} 

To compute the expectation value on the right hand side, we we expand the spinor field operator $\hat \psi$ in eigenfunctions of the Dirac Hamiltonian. The Dirac equation is $(\Dslash + m) \psi = 0$, and we make the following ansatz for the eigenfunction $\psi$:
\begin{equation} \label{tranS}
\psi =  z^{3/2} e^{-i \omega t + i \vec k \cdot \vec x}  \psis(z)\,.
\end{equation}  

For any fixed $\vec k$, the Dirac equation can be block-diagonalized \cite{Faulkner:2009wj}. Without loss of generality we set $\vec k = k \hat x$, and we let $\psis = \left( \psis_{+}, \psis_{-} \right)^t $. Using the following basis for the Clifford algebra:
\begin{equation}
\label{eq:realbasis}
\Gamma^{\uz} =   \sigma^3 \otimes \Ione , ~
\Gamma^{\ut} =  i  \sigma^1 \otimes \Ione , ~
\Gamma^{\ux} =  \sigma^2 \otimes \sigma^3 , ~
\Gamma^{\uy} =  \sigma^2 \otimes \sigma^1,
\end{equation} 
we have
\begin{equation}\label{dirac equation}
 \left[i\sigma^2 \partial_z - \sigma^2 \frac{m}{z} \pm k \sigma^3 + \Phi(z)\right] \psis_{\pm} = \omega \psis_{\pm}\,.
\end{equation}  

Eq. \eqref{dirac equation} has two linearly independent solutions, whose asymptotic behavior near the AdS boundary is
\begin{equation}
\label{eq:UVbc}               
\yandz_{\pm} \buildrel{z \to 0}\over {\sim}
a z^{-mL} \left( \begin{matrix} 0 \cr  1 \end{matrix}\right) 
+ b z^{mL} \left( \begin{matrix}  1  \cr 0 \end{matrix}\right)
\end{equation} 
plus terms subleading in $z$. We demand that the non-normalizable solution be zero\footnote {Allowing the mass to range over $(-\half, \infty)$, this includes the alternative quantization.}: $a = 0$. Moreover, as in \cite{Sachdev:2011ze}, we impose hard-wall boundary conditions at an IR cutoff $z=z_m$: 
the upper component of both $\psis_\pm(z_m)$ must vanish.
 
With these boundary conditions, the Dirac Hamiltonian is a self-adjoint differential operator with spectrum $\omega_{n, \vec k, s}$, labelled by a discrete index $n$, by the  momentum $\vec k$ and by the sign $s = \pm$ that distinguishes the upper components from the lower components. We denote the eigenfunctions with $\psis_{n,\vec k, s}$.

In order to give definite meaning to the expectation value in \eqref{gauss law 1}, for a given profile of of $\Phi$, we fill all the states with $\omega_{n, k, s} < 0$, and we subtract the same expectation value for $\Phi = 0$. That is, we solve
\begin{equation} \label{eq:gauss}
- \Phi''(z) = q^2 \left[n(z)|_\Phi - n(z)|_{\Phi = 0}\right] \equiv q^2\Delta n(z)\,,
\end{equation} 
with
\begin{equation} \label{number density}
 n(z) = \sum_{\vec k,n,s} \theta(-\omega_{\vec k,n,s})\psis_{n,\vec k,s}^\dag(z) \psis_{n, \vec k,s}(z)\,.
\end{equation} 

Clearly \eqref{number density} involves two sums that need to be regulated. We regulate the sum over $n$ by discretizing the $z$ coordinate, with lattice spacing $\Delta z$, and we impose a hard cutoff ${\vec k}^2 < \Lambda_k^2$ on the sum over momenta. After the subtraction in \eqref{eq:gauss}, and appropriate renormalization of the charge $q$, the problem has a well-defined limit as $\Delta z \to 0$, $\Lambda_k \to \infty$. Additional information on the subtraction and renormalization can be found in the supplementary material.

Eq \eqref{eq:gauss} also needs to be complemented with appropriate boundary conditions on $\Phi$. We want a finite chemical potential in the boundary theory, so we set $\Phi(0) = -\mu$, and we also impose $\Phi'(z_m) = 0$. We expect that the boundary condition at $z_m$ becomes unimportant as $z_m \to \infty$, and we verified this by exploring mixed boundary conditions as well, without finding a significant influence in the interior, for large enough $z_m$.

We solve the integro-differential system formed by \eqref{dirac equation} and \eqref{eq:gauss} by an iterative method, whereby the number density computed with a given a profile of $\Phi$ is used to update $\Phi$ through \eqref{eq:gauss} and the new profile is used to update the number density, until convergence.
\section{The resulting groundstate}

\begin{figure}[h] \begin{center}
\hskip-.2in
\includegraphics[width=250pt]{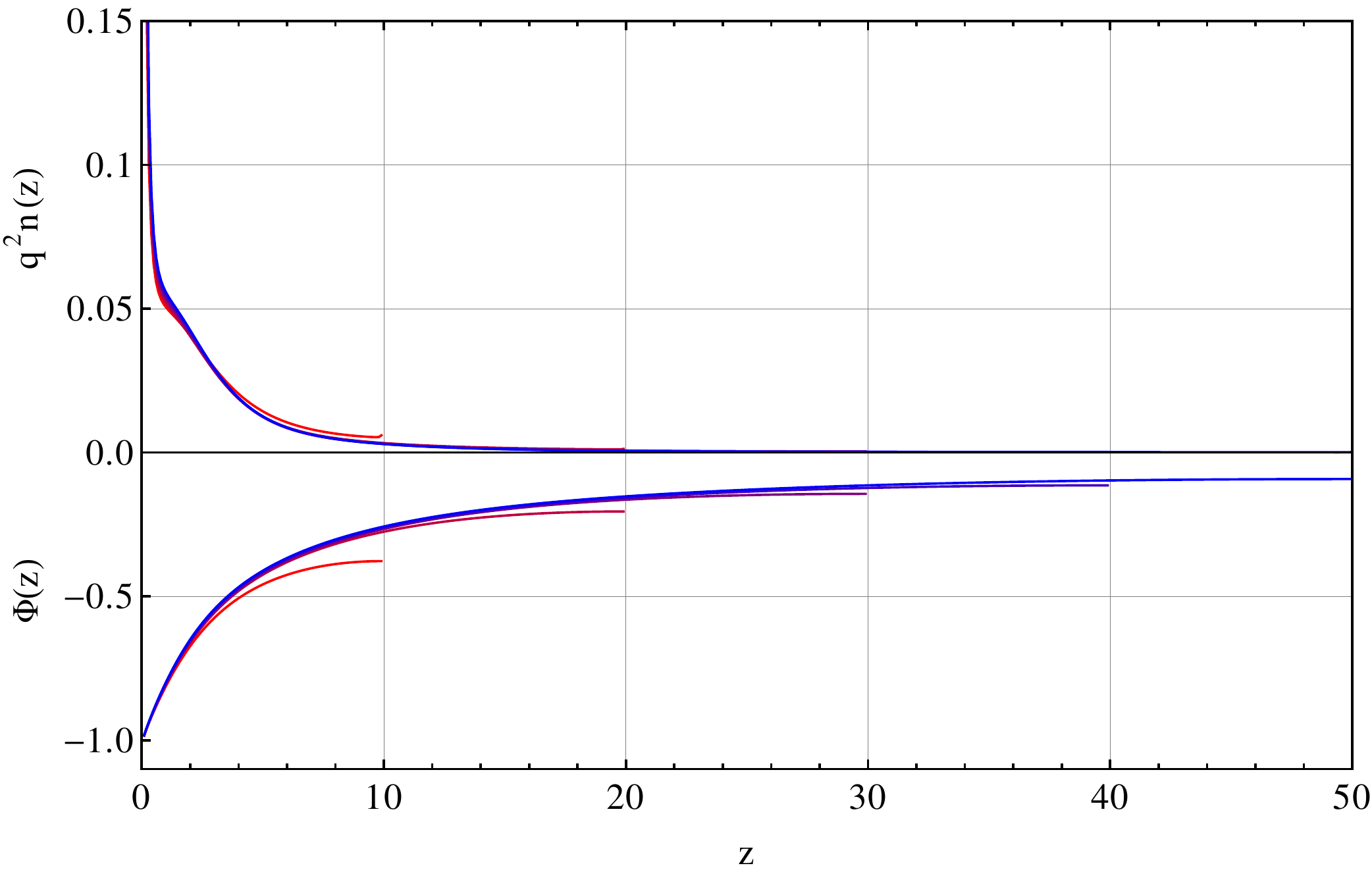}
\end{center}
\caption{
\label{fig:limit-exists}
This plot demonstrates the existence of the limit $z_m\to \infty$ with, from red to blue, $z_m = 10,\, 20,\, 30,\, 40,\, 50$. The associated values of the boundary chage density are
$\rho = 0.1710,\,0.1669,\,0.1638,\,0.1627,\,0.1622$.
$m = 0.3$;
$q = 2.0$;
$dz = 0.1$;
$\Lambda_k = 20$.
}
\end{figure}

\begin{figure}[h] \begin{center}
\includegraphics[width=250pt]{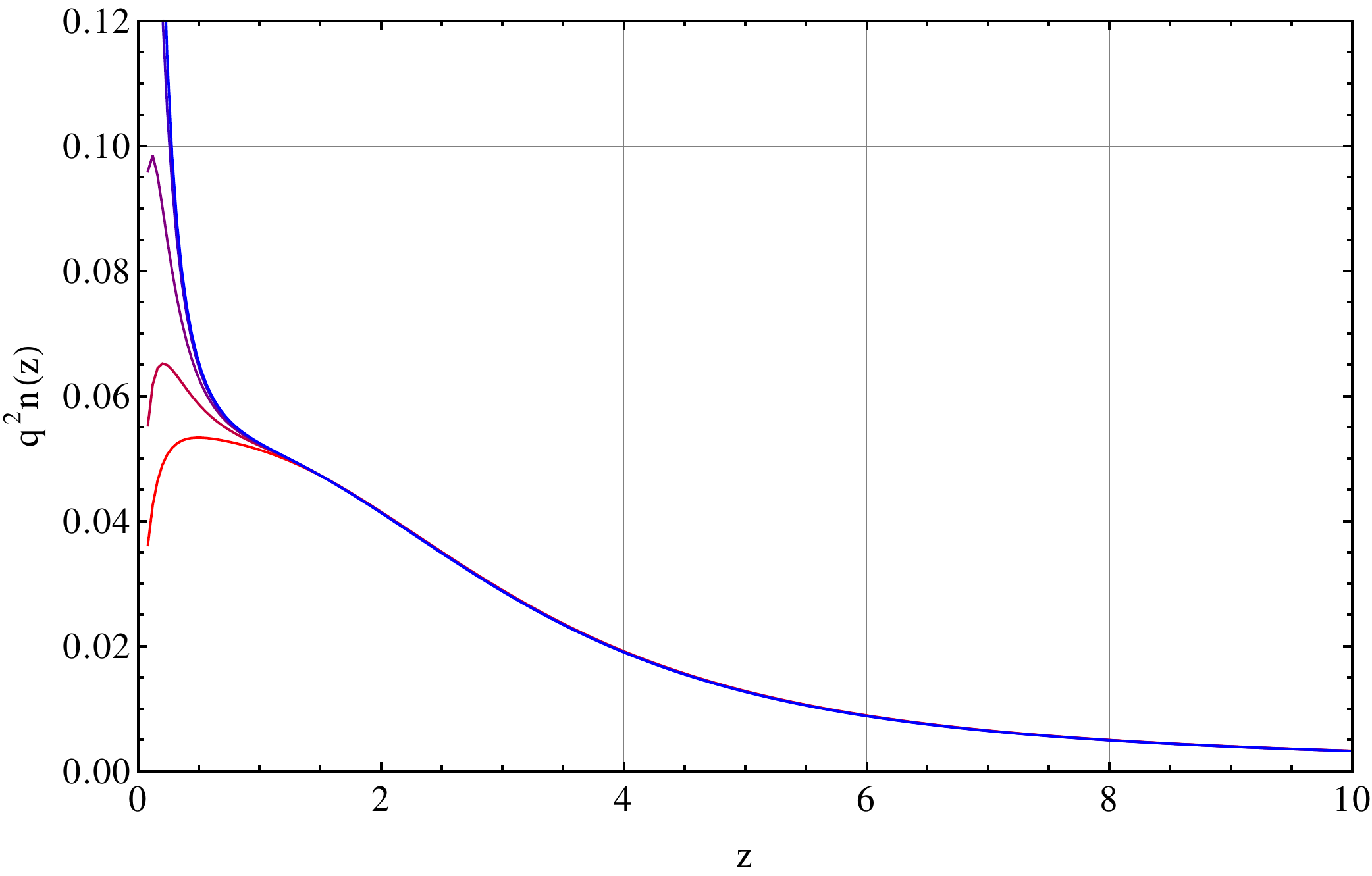}
\end{center}
\caption{
\label{fig:lambda-limit}
This plot displays how the limit $\Lambda_k \to \infty$ is approached, with, from red to blue, $\Lambda_k = 2,\, 3,\, 5,\, 10,\,20$. See the appendix for further discussion of the surface charge density. Corresponding values of the boundary charge density are $\rho = 0.1721,\, 0.1717,\, 0.1714,\, 0.1710,\, 0.1708$ (note that here $z_m = 20$, so these numbers should be compared with the first value in fig.~\ref{fig:limit-exists}). Positive $k$ represent the spectrum of the 
$m = 0.3$;
$q = 2.0$;
$z_m = 20$;
$dz = 0.04$.
}
\end{figure}

Fig.~\ref{fig:limit-exists} and \ref{fig:lambda-limit} display typical profiles for $\Phi(z)$ and $n(z)$. They 
also show that the problem possesses a well-defined limit as $z_m \to \infty$, $\Lambda_k \to \infty$. We also verified that the profiles are insensitive to the discretization of the $z$ coordinate, for small enough lattice spacing. Once the cutoffs are removed, the only remaining scale in the problem is the chemical potential $\mu$, and, without loss of generality, we can set $\mu = 1$. 

\begin{figure}[h] 
\begin{center}
\hskip-.2in
\includegraphics[height=250pt]{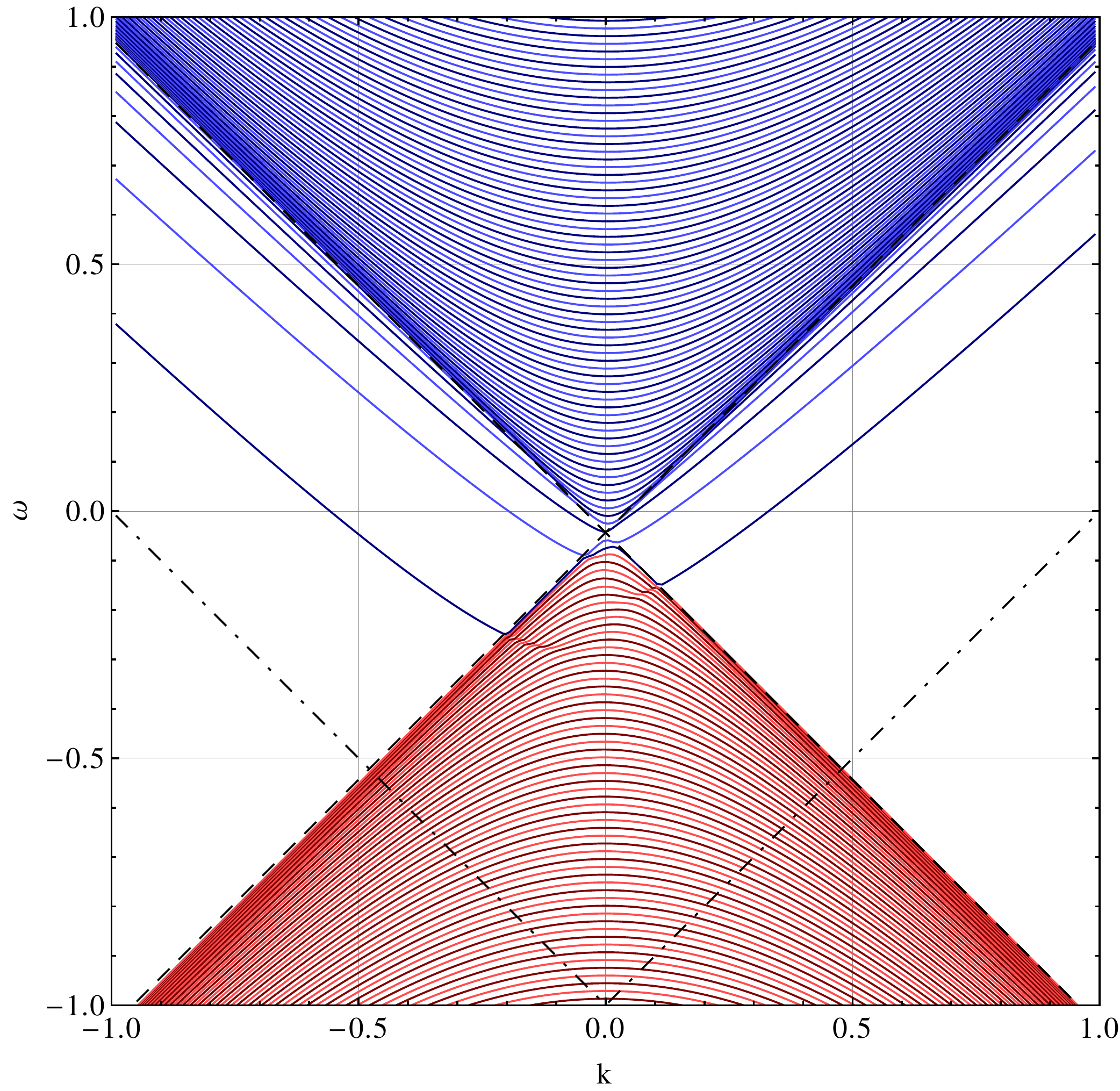}
\end{center}
\caption{\label{fig:spectrum}
The portion of the spectrum of the Dirac operator 
nearest to the chemical potential. For $k > 0$ we display $\omega_{n,k,+}$, for $k<0$ we display $\omega_{n,k,-}$. Also shown are the IR lighcone (dashed lines) and the UV lightcone (dash-dotted lines). It is clear from the figure that a continuum is developing inside the IR lightcone. $m = 0.3$; $q = 2.0$; $z_m = 200$; $dz = 0.2$; $\Lambda_k = 20$.
}
\end{figure}
The profile of the potential approaches a constant as $z \to \infty$\footnote
{If the electric field $\propto \Phi'$ does not vanish at the Poincar\'e horizon, 
an argument similar to \S7.4 of \cite{Hartnoll:2009ns} 
indicates that there will be a value of $z$ beyond which 
backreaction cannot be ignored, no matter how 
small the Newton's constant.}, and we would like to argue that this constant is zero.

The asymptotic behavior of the wavefunction for large $z$ depends on the sign of
\begin{equation}
 \label{IRlightcone} \kappa^2 \equiv (\omega - \Phi(\infty))^2 - k^2\,.
\end{equation} 
For $\kappa^2 < 0$ the wavefunction is exponentially decaying, whereas for $\kappa^2 > 0$,  $\psis(z) \sim e^{i \kappa z}$. Let us call the region $\kappa^2 > 0$ the infrared (IR) lightcone. As we increase $z_m$, the gap between the bands inside IR light cone decreases like $1 / z_m$. Hence, for $z_m \to \infty$, a continuum develops inside the IR light cone (fig.~\ref{fig:spectrum}). The contribution to the number density coming from each state within the IR light cone also decreases like $1 / z_m$, and is constant in $z$, for large $z$. Therefore, any finite portion of the lightcone that lies below $\omega = 0$ gives a finite, $z$-independent contribution to the number density.

If $\Phi(\infty) < 0$, after the subtraction of \eqref{eq:gauss}, the lower half of the light cone does not contribute any net number density, but the portion of the upper half of the light cone that lies below $\omega = 0$ gives, at large $z$, a finite, $z$-independent contribution to the number density. This is incompatible with the potential going to a constant as $z\to \infty$, due to Gauss law. A similar argument obtains for the case $\Phi(\infty) > 0$. The only possibility is for the potential to go to zero, so that the states within the IR light cone do not contribute at all to the number density density. Then, the only contribution comes from the few bands that lie outside the IR light cone. The corresponding wavefunctions are exponentially decaying, they don't contribute to the number density at large $z$, and hence they are compatible with the potential approaching a constant (zero) as $z \to \infty$. 

Determining numerically the exact nature of the falloff of the number density at large $z$ is very difficult, but it is quite manifestly subexponential. A possible explanation for this is that the bands that lie outside of the IR light cone, for a certain range of $k$, skirt the edge of the cone (follow the dashed line), as can be seen in fig.~\ref{fig:spectrum}. When the state is close to the edge of the cone, the rate of decay is very weak. Since the distance from the edge is a decreasing function of $z_m$, it is conceivable to obtain a a subexponential decay of the number density as $z_m \to \infty$.

From the holographic point of view, the IR region of the geometry is dual to a relativistic CFT, which only has spectral weight inside the lightcone $ |\omega| <  c k $. One interesting quantity that can be extracted from our computation is the boundary charge density $\rho$. This is the response to the chemical potential, and it is given by
\begin{equation}\label{boundary charge density}
 \rho = \Phi'(0) = q^2 \int \dr z\ \Delta n(z)\,,
\end{equation} 
where the second equality is a consequence of Gauss law. 
The charge density in the boundary is equal to the total charge in the bulk. Since $\mu$ is the only scale in the problem, its dependence on $\mu$ is determined by dimensional analysis to be $ \rho = A \mu^2 $, for some constant $A(q,m)$.

Let us also point out that \eqref{boundary charge density} guarantees Luttinger's theorem in the 
boundary \cite{Sachdev:2011ze}. Luttinger's theorem states that, for interacting fermions, the area of the Fermi surface is proportional to the number density, with the same factor as for free fermions. As will be discussed in more detail in the next section, there is a Fermi surface wherever one of the bands in fig.~\ref{fig:spectrum} crosses $\omega = 0$. According to our construction of $\Delta n(z)$, each $k$ mode within the Fermi surface contributes 
1 to integral on the RHS of \eqref{boundary charge density}, thereby ensuring Luttinger's theorem.

\section{Green's functions}

\begin{figure}[h] \begin{center}
\hskip-.2in
\includegraphics[width=250pt]{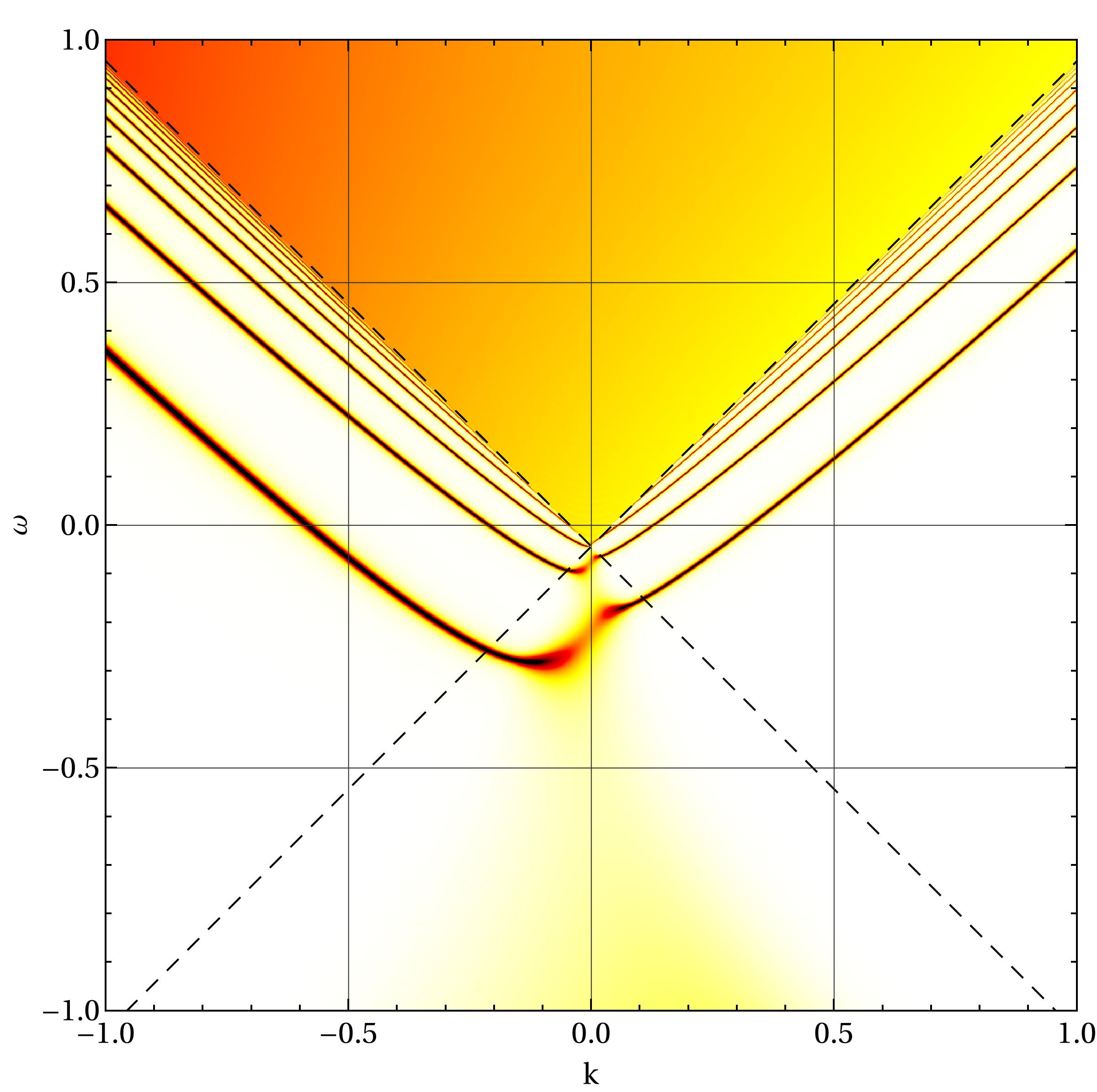}
\end{center}
\caption{
\label{fig:gr}
Density plot of the spectral density $\Im G_R(\omega, k)$, displaying the spectra of the stable quasiparticles, which coincide with the bands in fig.~\ref{fig:spectrum}, and the continuum inside the IR light cone. Notice the increased width of the quasiparticle peaks as they enter the light cone.
$m = 0.3$;
$q = 2.0$;
$z_m = 200$;
$dz = 0.2$;
$\Lambda_k = 20$.
}
\end{figure}

We compute the photoemission response of our state, proportional to the single-fermion spectral density $\Im G_R(\omega, k)$, where $G_R$  is the retarded single-fermion Green's function. To compute the retarded function, we impose in-falling boundary condition for the Dirac field at the Poincar\'e horizon, and include a source at the UV boundary in \eqref{eq:UVbc}, $a \neq 0$.  Then $G_R(\omega, k) = b/a$.  See fig.~\ref{fig:gr}.

The lifetimes of quasiparticles in holographic Fermi surfaces can be understood in terms of interactions with other gapless degrees of freedom \cite{Faulkner:2009wj, Faulkner:2009am, Faulkner:2010tq, Faulkner:2010zz}.
The degrees of freedom into which a Fermi surface quasiparticle might decay are those inside the IR lightcone 
described above.

Outside of the IR light cone, for $\omega = \omega_{n, k}$, there exists a solution to the Dirac equation that is real, decays exponentially at large $z$ and is normalizable in the UV. This means that the infalling boundary condition is trivially satisfied, because the wavefunction is zero at the horizon, and we have a finite response $b\neq 0$ for zero $a$. As a consequence, $G_R(\omega, k)$ has a pole at $\omega = \omega_{n, k}$, and hence there is a delta function singularity in $\Im G_R$. This delta function implies the existence of an exactly stable quasiparticle in the boundary theory. In particular, the points where $\omega_{n, k} = 0$ are Fermi surfaces with stable excitations. To detect such infinitely-narrow resonances in the numerics, we add a small imaginary part to the frequency, so as to move the pole away from the $\Re(\omega)$ axis, and convert the delta function to a narrow Lorentzian.

On the other hand, inside the IR lightcone, the asymptotic behavior of the wavefunction is $\psis \sim e^{\pm i\kappa}$, and hence, in general, a finite and complex $G_R(\omega, k)$ is needed to satisfy infalling boundary conditions. Consequently, quasiparticle excitations have finite width, because they can decay into 
the gapless CFT excitations. This phenomenon is visible in fig.~\ref{fig:gr} where the bound state bands enter the IR lightcone.

\section{Discussion}

The state we have discussed can be described semi-holographically \cite{Faulkner:2009wj, Faulkner:2010tq} as arising from a Fermi liquid coupled to a relativistic CFT. 
The study of Fermi surfaces coupled to critical systems has a long history, \eg~\cite{PhysRevB.14.1165, moriya, PhysRevB.48.7183, RevModPhys.79.1015, PhysRevB.82.075128, PhysRevB.82.075127, PhysRevB.82.045121, PhysRevB.84.125115}.
The conclusion of our holographic calculation is that the coupling between these sectors 
described here is an irrelevant deformation of the Landau theory.  In fact, according to our discussion about the stability of the quasiparticles, and the location of the IR light cone, the only possible singularities at $\omega = 0$, $k \neq 0$ are delta function peaks, which indicate exactly stable quasiparticles. This fact is very likely a consequence of the probe limit $G_N \to 0$.

The nature of the coupling between the FS and the CFT that one infers 
for the semi-holographic picture is a hybridization between a fermionic
operator of the CFT and the electron operator, as in \cite{Faulkner:2009wj, Faulkner:2010tq}.
Possibly-relevant couplings between the fermion density and relevant bosonic operators
of the CFT of the kind considered in 
\cite{PhysRevB.66.235117}
are suppressed in our large-$N$ limit.

It will be very interesting to study the effect of the screening by the fermions on the boundary gauge theory dynamics. On general grounds we expect that, with enough fermions, even beginning with a confining solution at $\mu = 0$, the gauge theory will deconfine. Holographically, this requires taking into account the gravitational backreaction of the bulk fermions. Progress in this direction will be reported elsewhere.
Resolving the problem confronted in this paper -- 
the question of the state of the bulk fermions in the presence of a horizon -- 
was an essential prior step.

\vskip.2in
{\bf Acknowledgements}
We thank T.~Faulkner, S.~Hartnoll, N.~Iqbal, S.~Sachdev, T.~Senthil, E.~Silverstein, 
B.~Swingle, D.~Vegh 
and the other participants 
at the KITP workshop ``Holographic duality and condensed matter physics"
for discussions, comments and encouragement.
We thank the KITP for hospitality.
This work was supported in part by
funds provided by the U.S. Department of Energy
(D.O.E.) under cooperative research agreement DE-FG0205ER41360,
in part by the Alfred P. Sloan Foundation,
and in part by the National Science Foundation under Grant No. NSF PHY05-51164.
Simulations were done on the MIT LNS Tier 2 cluster,
using the Armadillo C++ library.

\bibliography{fl}


\section{Supplementary material}

\subsection{Renormalization prescription}

Here we describe our method to obtain a 
number density 
which is independent of the cutoffs $\Lambda_k$ and $\Delta z$, 
and whose $z$ integral is also cutoff-independent.

Since the ultraviolet divergences probe short-distance structure of the theory, we can understand them while working in the flat-space limit, 
\ie~to lowest order in an expansion in curvature. The expectation value of the fermion number current in a given background $A_\mu$ can be evaluated in perturbation theory
 as

 $\langle \bar \psi \gamma^\mu \psi \rangle = $ \parbox{0.45in}{\includegraphics[width=0.45in]{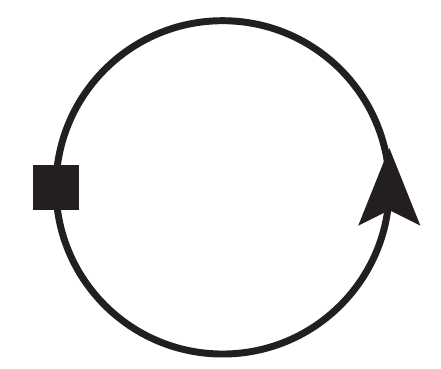}} +  \parbox{0.7in}{\includegraphics[width=0.7in]{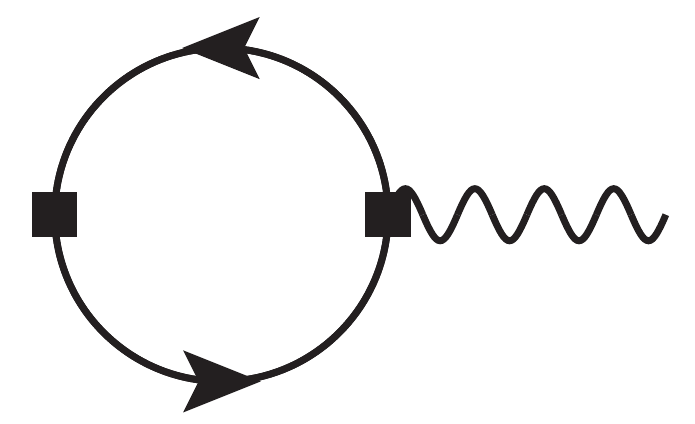}} + \parbox{0.7in}{
\includegraphics[width=0.7in]{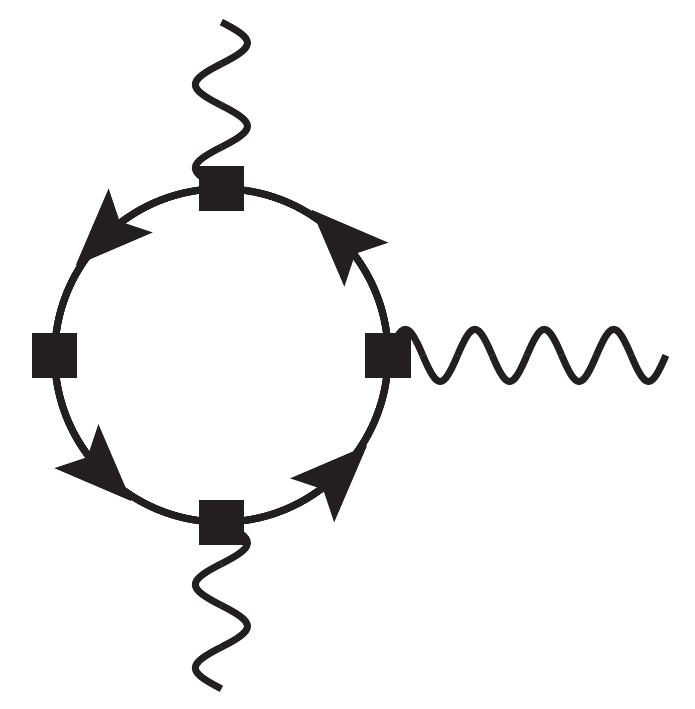}} + ... ,

\noindent where the box represents an insertion of the current,
and the wiggly lines represent the background field. 
Counting powers of loop momenta, 
diagrams with up to four insertions may be divergent. 
The first term is made finite by the subtraction of the zero-field value as in \eqref{eq:gauss};
this ensures that the density in vacuum is zero: 
$\Delta n_B \equiv \vev{\psi^\dagger \psi}_A - \vev{\psi^\dagger \psi}_{A=0}$.
Following familiar analysis from QED in flat space, 
loops with odd numbers of insertions vanish in the UV limit by charge conjugation invariance,
and the divergence in a loop with four insertions is ruled out by the Ward identity. 
In the remaining loop with two insertions, the naively expected quadratic divergence is again ruled out by the Ward identity, and the only remaining divergence is logarithmic. 
Specifically, the 
bare
density 
for a background field of the form $ A = \Phi(z) dt$ 
is
\begin{equation}
\begin{split}
\Delta n_B(z) &= {\delta W[A = \Phi dt] \over \delta \Phi(z)} \\
& =
\int \frac{\dr^4p}{(2\pi)^4}\tilde \Phi(p)e^{-i p_z z} \times \\
& \times  \int \frac{\dr^4k}{(2\pi)^4}\tr[\gamma^0 \tilde D_F(k)\gamma^0\tilde D_{F}(k+p)] + \mathcal O(\Phi^2)\label{eq:inducedcd2}\\
&=
\frac{4}{3}
\left[\int \frac{\dr^4l_E}{(2\pi)^4}\frac{1}{l_E^4}\right]\:\Phi''(z)+ \text{finite terms}
\end{split}
\end{equation}
where $\tilde \Phi(p)$ has delta-function support at $p_t=p_x=p_y=0$, $\tilde D_F$ is the momentum-space fermion propagator, and the subscript B denotes a bare quantity. 

Thus, with a Lorentz-invariant cutoff $\Lambda$, the logarithmic divergence in the bare density is 
\begin{equation}
 \label{eq:log} \Delta n_B(z) =  \frac{1}{6\pi^2} \ln \frac{\Lambda}{M}\, \Phi''(z) +\ \text{finite terms}\,,
\end{equation}
where $M$ denotes some infrared energy scale.
In our non-relativistic numerical setup with two cutoff scales $\Lambda_k$, $1/\Delta z$, the divergence has a more complicated form 
\begin{equation}\label{susceptibility}
\Delta n_B(z) =  
\chi(\Lambda_k,\Delta z)\,\Phi''(z) +\ \text{finite terms}\,, 
\end{equation} 
where $\chi$ diverges logarithmically with $s$ 
under the simultaneous scaling
 $\Lambda_k \to s \Lambda_k$, $\Delta z \to \Delta z /s$. 
 In the next subsection we 
use perturbation theory to examine 
in more detail the number density and the function $\chi$, in the limit $\Delta z \to 0$, $\Lambda_k$ finite. 

In the numerical analysis, we compute $\chi(\Lambda_k, \Delta z)$ 
at a given value of the cutoffs from the charge response to a small background field of conventional profile \mbox{$\Phi_\star(z) = \varepsilon z^2$}.  
In order to be able to neglect the finite, nonlinear terms in the response \eqref{susceptibility}, 
we require $ \varepsilon \ll z_m^{-3}$; 
the quadratic profile conveniently removes any higher-derivative contributions.

Once $\chi$ is known, for any given $\Phi$, we define a renormalized number density as 
\begin{equation}\label{renormalized n}
\Delta n_R(z) \equiv \Delta n_B(z) - \chi(\Lambda_k,\Delta z) \left[\Phi''(z) + \delta(z) \Phi'(z)\right]~.
\end{equation}
In the iteration procedure, 
the electrostatic potential for the next step
$\Phi_\text{next}$ 
is the potential produced by the renormalized density according to
\begin{equation}\label{renormalized gauss law}
 -\Phi_\text{next}''(z) = q_R^2 \Delta n_R(z)~.
\end{equation} 
The renormalized charge $q_R$ is by definition cutoff-independent --
it is the charge parameter specified in the plots above.
Apart from the delta function term, the subtraction \eqref{renormalized n}
has the effect of renormalizing the charge $q$ according to
\begin{equation}
  q_R^2 =  \frac{q_B^2}{1 + q_B^2 \chi(\Lambda_k, \Delta z)}\,,
\end{equation} 
as can be readily seen by substituting this and \eqref{renormalized n} into \eqref{renormalized gauss law}. 
Away from the boundary, this implies $q_R^2 \Delta n_R=q_B^2 \Delta n_B$.

The delta-function term in \eqref{renormalized n} is required by the following logic.
Its purpose is to guarantee that integrating both sides of \eqref{renormalized n} over $z$ gives the relation
\begin{equation}
 \int \dr z\ \Delta n_R(z) = \int \dr z\ \Delta n_B(z)\,.
\end{equation} 
This is very important, because the right hand side does not depend on the cutoff, for large enough $\Lambda_k$ 
(larger than any Fermi momenta, so that varying $\Lambda_k$ does not change which states are filled relative to the vacuum), since the contribution of each additional $k$-mode to the integral is then exactly zero. On the left hand side, the renormalized charge density should also not depend on the cutoff, hence the need to add the delta-function term at the boundary.


\subsection{Single mode linear response}

In this section we compute at first order in perturbation theory the number density response of a single $k$-mode, in the limit $\Delta z \to 0$. This is important as a check on the numerics, and for the interpretation of the surface charge in the next section. We will assume translational invariance, so our results will be valid when $k \gg 1/z_m$, for $0 \ll z \ll z_m$. In this regime the bulk fermion mass has no effect, so we focus on $m=0$.

We define the susceptibility $\bar \chi$ as
\begin{equation}\label{k mode susceptibility}
 \Delta n_k(z) = \int dz' \bar \chi(z, z') \Phi(z')\,,
\end{equation} 
and, at first order, we have: 
\begin{equation}
\label{eq:linear-response}
 \bar \chi(z, z') = \sum_{\scriptstyle m | \omega_m >0}\sum_{\scriptstyle n| \omega_n <0}
\frac{\psis^\dag_n(z)\psis_m(z)\psis_m(z')^\dag\psis_n(z')  }{\omega_n - \omega_m}
+ \text{h.c.},
\end{equation} 
where $\psis_m$ labels a basis of eigenstates of the unperturbed single particle Hamiltonian  $h = k \sigma^3 - i \sigma^1 \partial_z$. The eigenmodes are 
\begin{equation}
 \psis_{\kappa s}(z) = {e^{i \kappa z } \over \sqrt{2\pi}} \psis_{\kappa s}\,,
~~~(k \sigma^3 + \kappa \sigma^1 ) \psis_{\kappa s} = s \omega_{\kappa} \psis_{\kappa s}\,,
\end{equation} 
with $\omega_{\kappa}\equiv \sqrt{\kappa^2 +k^2} $ and $s = \pm 1$.
The explicit eigenvectors are 
\begin{align}
& \psis_{\kappa+} =
{1\over \sqrt{ 2  \omega_\kappa ( \omega_\kappa - k)}}
\begin{pmatrix}
\kappa 
\cr
\omega_{\kappa} - k 
\end{pmatrix}
\\
& \psis_{\kappa-} =
{1\over \sqrt{ 2  \omega_\kappa ( \omega_\kappa- k)}}
\begin{pmatrix}
\omega_{\kappa} - k \\
\kappa
\end{pmatrix}~.
\end{align}

Plugging these expressions into \eqref{eq:linear-response} gives
\begin{equation}
\bar \chi(z, z') = 
- \frac{k}{2\pi}\int \frac{\dr p}{2\pi} e^{i k p (z-z') } 
\Upsilon(p) + \text{h.c.} 
\end{equation} 
with 
\begin{equation}
 \label{eq:ups} \Upsilon(p) = 1 - \frac{4 \text{arctanh}\left( {p \over \sqrt{4 + p^2 }}\right) }{p \sqrt{4 + p^2 }}\,.
\end{equation} 

The fourier transform of $\Upsilon$ does not exist, but we can introduce a different susceptibility $\chi$ as
\begin{equation}
\Delta n_k(z) = \int \dr z' \chi(z-z') \Phi''(z')\,,
\end{equation} 
and we have a well defined answer
\begin{equation}
\chi(z) =  \int \frac{\dr \kappa}{2\pi} e^{i \kappa z } { \Upsilon\( \kappa/k\) \over 2\pi \kappa^2 }  + \text{h.c.}\,.
\end{equation} 

Using the explicit expression \eqref{eq:ups} for $\Upsilon$, we observe numerically that
\begin{equation}
\label{eq:susc-k}
\chi(z) \approx  {1\over a k} e^{- b |k z|}~, ~~~~~~a \approx 18.16, b \approx 2.17\,,
\end{equation} 
that is, the response of a mode of momentum $k$ is exponentially localized to a width
of order $ {1\over |k|}$ in the radial direction. 

\begin{figure}[h] \begin{center}
\hskip-0.2in
\includegraphics[width=250pt]{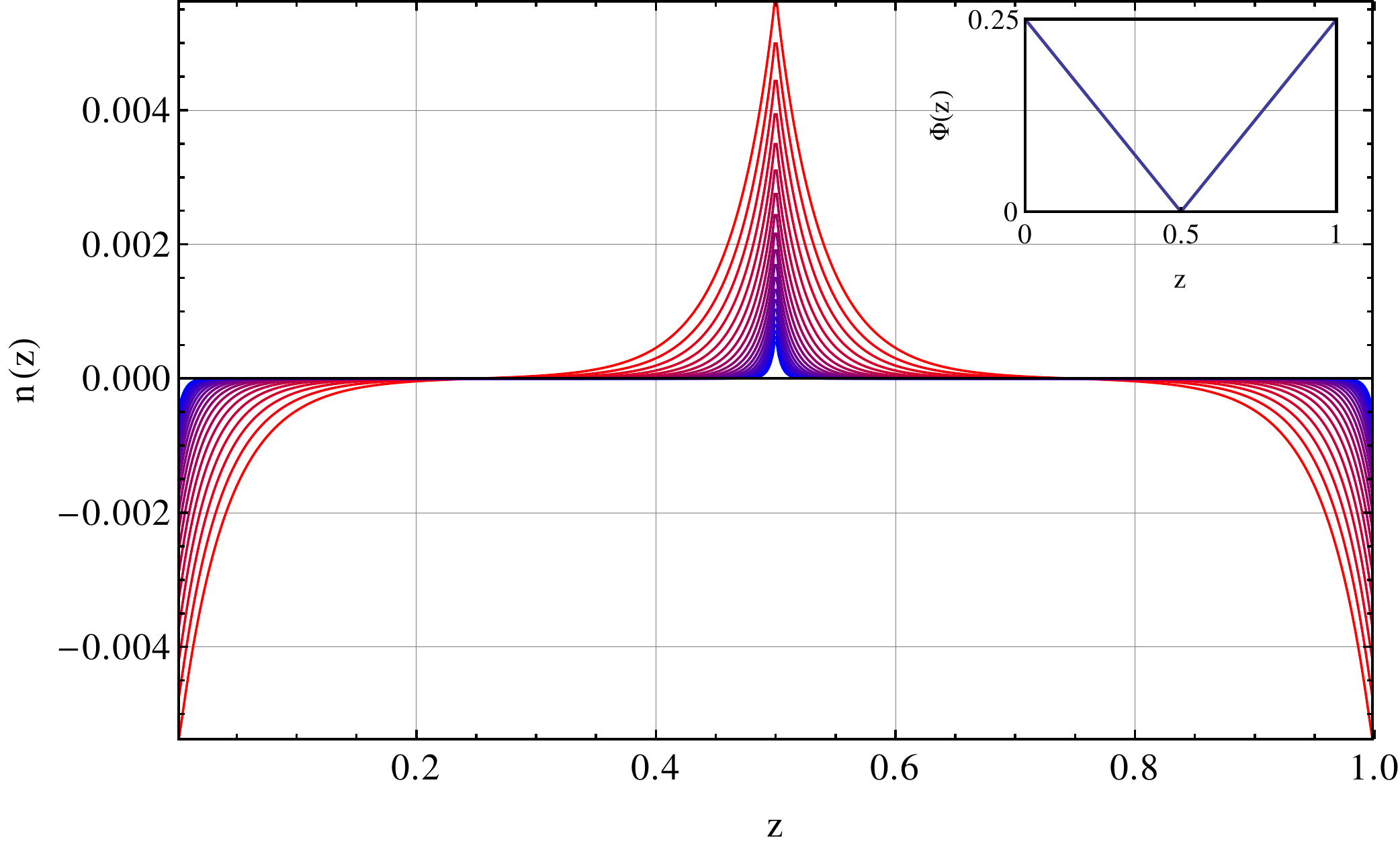}
\end{center}
\caption{
\label{fig:surface-charge}
In this figure we show the contribution to the charge density from modes with various momentum $k$, in response  to a profile of the electrostatic potential which has delta-function in the second derivative. The profile of $\Phi$ is indicated in the inset. Notice how the response to the source is similar to the finite size effects at the boundary. $k$ from 10 to 100 in logarithmic steps.
}
\end{figure}

We verified that the numerical computation of the number density, at large $k$, matches this result down to the coefficient $a$ and $b$. In particular, figure \ref{fig:surface-charge} shows the number density response to a source such that $\Phi''(z) = \delta(z - z_m /2)$.

In the special case of $\Phi_\star = \frac{1}{2} z^2$, which is (proportional to) the conventional profile we use to compute $\chi(\Lambda_k, \Delta z)$, we have
\begin{equation}
\Delta n_k = \int \dr z\ \chi(z) = {1\over \pi k^2 } \lim_{p \to 0} \frac{\Upsilon(p)}{p^2 } = \frac{1}{6 \pi k^2 }
\end{equation} 
Integrating this result over momenta reproduces the familiar logarithmic divergence \eqref{eq:log} ({\it q.v.}~\cite{Peskin:1995ev}, eqn (7.96)). More explicitly we have:
\begin{equation}
  \chi(\Lambda_k, \Delta z = 0) = 2 \int_M^{\Lambda_k} \frac{d^2k}{(2\pi)^2} \frac{1}{6\pi k^2} =  \frac{1}{6 \pi^2} \log \frac{\Lambda_k}{M}\,,
\end{equation} 
where $M$ is again some infrared scale, and the factor of 2 comes from the sum over spins.

\subsection{The surface charge}

\begin{figure}[h] 
\begin{center}
\includegraphics[width=250pt]{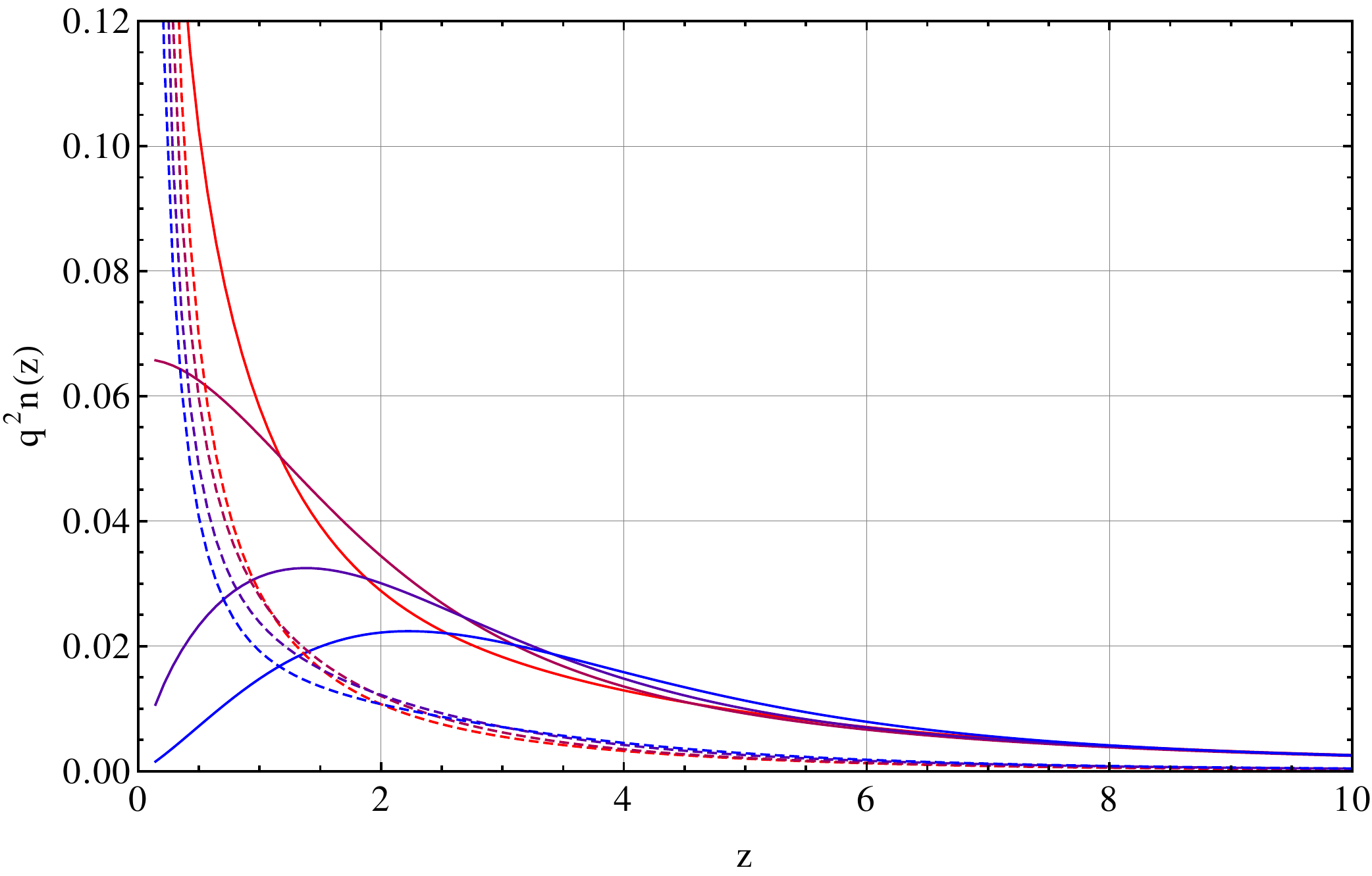}
\end{center}
\caption{
\label{fig:components}
This plot shows the separate contributions to the charge density from
the conduction bands (solid line) and valence bands (dashed line), with, from red to blue, $m = -0.3,\, 0.0,\, 0.3,\, 0.6$. The corresponding values of the boundary charge density are $\rho = 0.3105, 0.2165, 0.1623, 0.1304$. $q = 2.0$; $z_m = 50$; $dz = 0.071$; $\Lambda_k = 20$.
}
\end{figure}

The bulk system we are studying can effectively be compared to a metal. It possesses an (infinite) number of completely filled bands, which we will call valence bands, and a number of partially filled bands, which we will call conduction bands. If a transverse electric field $E_i$ is applied, there will be a current response. However, the boundary conditions prevent the charge from moving in the $z$ direction, like in an isolated slab of metal. Therefore, the fermions respond to the electric field $E_z = -\Phi'(z) < 0$ by attempting to screen it, in a way that can be compared to the polarization of an insulator. This is the origin of the thin layer of charge that is visible close to the AdS boundary in figs.~\ref{fig:limit-exists},\ref{fig:lambda-limit},\ref{fig:surface-charge},\ref{fig:components}.

The contributions to the layer from the valence and the conduction bands is very different, as can be seen in fig.~\ref{fig:components}. In particular, the contribution from the valence bands forms a thin layer, which is largely independent of the fermion mass; we infer that it is sustained by wavefunction gradient pressure, \ie~the uncertainty principle. On the other hand, the contribution from the conduction bands depends strongly on the fermion mass, and a positive mass has the effect of pushing away the charge density from the boundary.

From the point of view of the field theory dual of our bulk system, this surface charge at the boundary of $AdS$ can be interpreted as the screening due to the rearrangement of the short-wavelength modes of the CFT.

\begin{figure}[h] \begin{center}
\includegraphics[width=250pt]{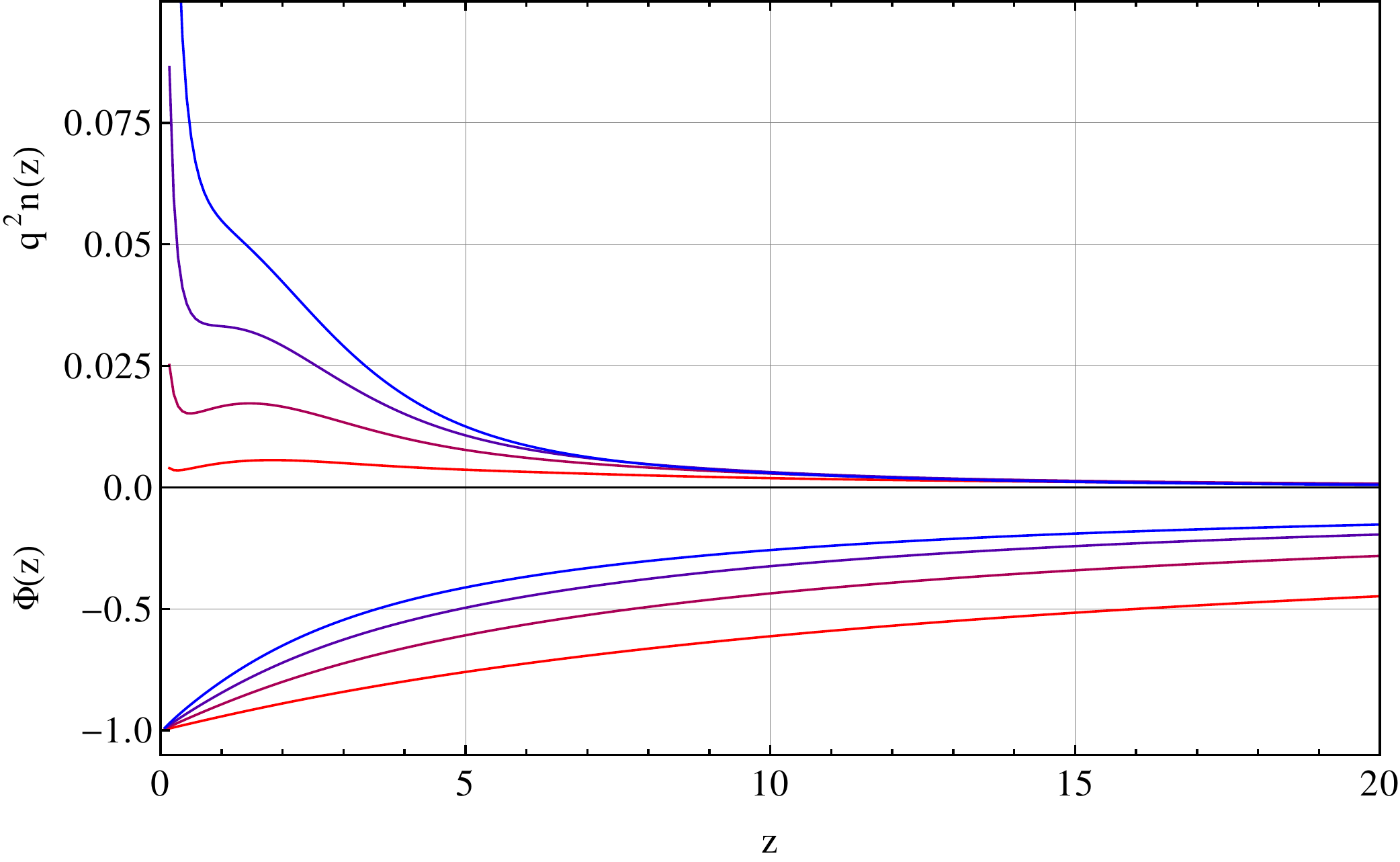}
\end{center}
\caption{
\label{fig:charge}
This plot shows the dependence on $q$, with, from red to blue, $q = 0.5,\,1.0,\, 1.5,\, 2.0$. The corresponding values of the total charge are $\rho = 0.0584,0.1055,0.1411,0.1623$.
$ m = 0.3$;
$z_m = 50$;
$dz = 0.071$;
$\Lambda_k = 20$.
}
\end{figure}

Fig.~\ref{fig:charge} shows the effect of the charge on the total number density and on the profile of the field. A larger charge leads to a more rapid screening of the chemical potential. The surface charge also becomes comparatively larger, due to the stronger electric field at the boundary.

\subsection{The bulk chiral anomaly}

An important check of our regularization of the Dirac Hamiltonian is the ability to reproduce the chiral anomaly, in the following way. 

Using the Dirac equation in the continuum, one can show that, for each mode $\psis_{n, k, s}\equiv \psis$ and for any choice of $\Phi$, the following equation holds:
\be
\partial_z \(\psis^\dagger \psis\) = 2 \psis^\dagger \(-m \sigma^3 \pm 
k \sigma^1\) \psis \,.
\ee
A naive application of this statement would tell us that, for $m=0$ and $k=0$, the profile of the charge density 
is a constant in $z$:
\be\label{classical-anomaly} 
\partial_z n_k  \buildrel{?}\over{=} 0
\quad \text{for} \quad m = k = 0\,.
\ee

This statement is in fact false.
Indeed, the unregulated statement \eqref{classical-anomaly} asserts the conservation, at the classical level, of 
an axial current. This conservation law is violated quantum mechanically,
in the presence of a background gauge field coupling to the vector-like current. 
In fact, if we limit ourselves to study the $k=0$ mode, the bulk Dirac problem is that of a 1+1 dimensional fermion (or rather two such fermions, because the spin index becomes a species index), described by the action
\begin{equation}
 S = \int \dr z\, \dr t\ 
 i \bar \psis ( \Dslash + m ) \psis\,
\end{equation}
with $ \Dslash \equiv \gamma^z \partial_z + \gamma^t \( \partial_t + i \Phi \) $.

When $m=0$, this action is invariant under chiral rotations $\psis \mapsto e^{i \alpha \gamma^5} \psis$, and hence the axial current $j_5^\mu = \bar \psis \gamma^5 \gamma^\mu \psis$ is classically conserved. If we define $\gamma^5 \equiv \gamma^z \gamma^t $, we have $j_5^z = \psis^\dag \psis$, and hence \eqref{classical-anomaly} states the conservation of the axial current.

At the quantum level, the axial current conservation is violated by the chiral anomaly:
\be 
\partial_\mu j_5^\mu = {1 \over 2\pi}\epsilon_{\mu\nu}F^{\mu\nu}\,.
\ee
If we integrate this equation with respect to $z$ we get 
\be 
n_k(z) = - {1\over \pi} \Phi + \text{const}
\quad \text{for} \quad m = k = 0\,,
\ee
which we observe numerically.


\end{document}